\documentstyle[12pt]{article}

\makeatletter
\@addtoreset{equation}{section}
\makeatother

\begin{document}
\hfill{}
\hfill{}

\hfill{UCSBTH-96-27}

\hfill{hep-th/9610170}

\vspace{24pt}

\begin{center}
{\large {\bf Composite black holes in external fields}}

\vspace{48pt}

Roberto Emparan

\vspace{12pt}

{\sl Department of Physics}\\
{\sl University of California}\\
{\sl Santa Barbara, CA 93106}\\
{\it emparan@cosmic1.physics.ucsb.edu}

\vspace{72pt}

{\bf Abstract}
\end{center}
\begin{quote}{
The properties of composite black holes in the background of electric
or magnetic flux tubes are analyzed, both when the black holes remain
in static equilibrium and when they accelerate under a net external
force. To this effect, we present a number of exact solutions
(generalizing the Melvin, C and Ernst solutions) describing these
configurations in a theory that admits composite black holes with an
arbitrary number of constituents. The compositeness property is argued to
be independent of supersymmetry. Even if, in general, the shape of the
horizon is distorted by the fields, the dependence of the extreme
black hole area on the charges is shown to remain unchanged by either
the external fields or the acceleration. We also discuss pair creation
of composite black holes. In particular, we extend a previous analysis
of pair creation of massless holes. Finally, we give the generalization
of our solutions to include non-extreme black holes.
}
\end{quote}

\newpage

\section{Introduction}

The recently successful microscopic interpretation
of the Bekenstein-Hawking entropy within string theory \cite{it}
has required, among other things, a deeper understanding
of the black hole solutions of the low energy field equations of the
theory.
In particular, an important role has been played by a class of
remarkably simple extremal black hole solutions with charges
belonging to different gauge fields
\cite{cvetyoum2,cvetsey,rahmfeld,comptse}. The way in which these charges
enter in the solutions, namely, through products of harmonic functions,
indicates that each of the gauge fields acts independently of
the others. This feature, regardless of the different
stringy origin of each of the gauge fields,
leads to a picture in which
the extremal black hole can be viewed as a composite of
three (in five dimensions) or four (in four dimensions)
`constituent' black holes, each of the latter possesing charge
under only one of the gauge fields.

The simplest generalization of Einstein-Maxwell theory that is relevant
to superstring and supergravity theories is obtained by introducing
a scalar (dilaton) field $\phi$ whose coupling to the Maxwell field is
measured by some constant $a$. The action takes the form
\begin{equation}
\label{dilaction}
I={1\over 16\pi G}\int d^4x\sqrt{-g}\biggl\{ R-2(\partial\phi)^2
-e^{-2a\phi} F^2\biggr\}.
\end{equation}
The fields $F$ and $\phi$ are in general linear combinations of the
large variety of gauge fields and scalars
of the underlying theories. Black hole solutions of the equations of
motion of these theories were found in \cite{gibmae}. For
$a>0$, the event horizon of the black holes in the extremal limit
shrinks to zero size and becomes singular.
However, it was found in \cite{klop} that if one considers a theory with
two different gauge fields (and $a=1$), then the extremal horizon
is regular
as long as the charges under each $U(1)$ are both non vanishing.
These $U(1)^2$ theories
of \cite{klop} can in turn be embedded in the following $U(1)^4$
theory, which arises as a truncation of low energy heterotic string
theory
compactified on a six-torus \cite{rahmfeld}
\begin{eqnarray}
\label{hetaction}
I&=&{1\over 16\pi G}\int d^4x\sqrt{-g}\biggl\{ R-{1\over 2}\left[
(\partial\eta)^2 + (\partial\sigma)^2 +(\partial\rho)^2\right]\nonumber
\\
&-&{e^{-\eta}\over 4}\left[ e^{-\sigma-\rho}F_{(1)}^2
+ e^{-\sigma+\rho}F_{(2)}^2
+e^{\sigma+\rho}F_{(3)}^2 +e^{\sigma-\rho}F_{(4)}^2\right]\biggr\}.
\end{eqnarray}
The composite black hole solutions (to be reviewed below) admit also
different embeddings into string
theory other than toroidally compactified heterotic strings,
as well as in eleven-dimensional supergravity
(see, e.g., \cite{comptse} and
refs.\ therein).
This action, and its extreme black hole solutions, has been used
to argue that the extreme black holes
of (\ref{dilaction}) with $a=\sqrt{3},1,1/\sqrt{3},0$ consist
of, respectively, 1, 2, 3, or 4 particle states bound at threshold
\cite{rahmfeld,duffrahm2}. It should be noted that these particular
values of $a$ are the only
ones for which the black holes can be consistently embedded in
maximal supergravity theories.

We will also propose another class of actions, which generalize the
$U(1)^4$ theory above,
and which will be used to construct composites of an arbitrary number
of extreme black
holes. Specifically, the theories contain $n$ gauge fields and $n-1$
independent scalars with action
\begin{equation}
\label{unaction}
I={1\over 16\pi G}\int d^4x\sqrt{-g}\biggl\{ R-{1\over 2 n^2}
\sum_{i=1}^n\sum_{j=i+1}^n(\partial\sigma_i-\partial\sigma_j)^2
-{1\over n}\sum_{i=1}^n
e^{-\sigma_i} F_{(i)}^2\biggr\},
\end{equation}
and the scalars satisfying
\begin{equation}
\label{singular}
\sum_{i=1}^n\sigma_i =0.
\end{equation}
This theory
contains, modulo electric-magnetic duality rotations of the
gauge fields, the $U(1)^2$
theories of \cite{klop} and the $U(1)^4$ theory above
\footnote{The theory
(\ref{unaction}) is closely related to the multi-scalar theories
considered in \cite{lupope}. However, the constraint (\ref{singular})
that we are imposing implies that we consider, in general, their matrix
$M_{\alpha\beta}$ to be singular. The only
such case considered in \cite{lupope}
in 4 dimensions is the $U(1)^4$ theory, embeddable
in a maximal supergravity theory.}. The extremal black hole solutions
of the single-scalar-Maxwell theory
with any rational value of $a^2$ will be seen to arise as composites
of several of these black holes. In all cases the binding energy will
be zero.

The action (\ref{unaction}) does not, in general, arise from any
truncation of compactified string theory, but it might instead appear
from some truncated form of extended supergravity coupled to
matter supermultiplets. Though we have not found such embedding,
the extremal black hole solutions of (\ref{unaction})
present a number of features in common with
supersymmetric black holes. These include saturation of BPS-like bounds,
multi-center solutions and attractors in moduli space. The solutions
are found to admit simple forms even for an arbitrary number of gauge
fields, and this will allow a study in wide generality of
the properties of composite black holes.

The main goal of this paper will be to probe the behavior
of composite black holes in external fields.
To do so, we have to analyze how to describe the geometry of these black
holes in an external field.
In a theory containing gravity a uniform electric or
magnetic field does not exist: the energy stored in the field bends the
spacetime, and the most we can aspire to is an axisymmetric configuration
describing a flux tube with field strength approximately uniform on
distances to the axis smaller
than the magnetic length $1/B$. This is the Melvin
solution of General
Relativity \cite{melvin}.
As will be shown below, the theories (\ref{hetaction}) and
(\ref{unaction}) also admit Melvin flux tube solutions,
with the fields taking
the form of composites.

What can we expect to happen to a composite black hole when placed
in a
background field? First of all, we have each of the $U(1)$ charges $q_i$
coupled to a background flux tube of strength $B_i$, and thus
subject to a
force given (roughly) by $q_i B_i$.\footnote{The charges and fields can
be either electric or magnetic, but we will indistinctly denote them
by $q_i$ and $B_i$.}
We can therefore achieve an equilibrium configuration for the
black hole if the forces are balanced so that their total
cancels out. One may wonder whether the
composite should
not break apart, given the zero binding energy between constituents.
This question turns out to be not easy to ascertain.
In the solutions of black holes in external fields that we will
present, the constituents
remain bound almost by construction.
The problem is that
the supersymmetry of the extreme solutions without external fields
that guarantees their stability and the cancellation of forces
between constituents, is completely broken
by the introduction of the background field, which does not possess any
Killing spinors.
We envisage two possibilities:

a) The solution describing the composite black hole in
equilibrium under
the pull of different $U(1)$ fields could be classically unstable,
evolving perhaps towards a configuration where the composite is
broken. The analysis of this possibility, that involves studying the
spectrum of fluctuations of
the solution, is certainly extremely
complicated due to the low degree of symmetry of the solution.
Even more, the stability analysis could change if we embedded the
black hole in a higher dimensional theory. However, the classical
black hole theorems forbid the possibility of classical splitting of
the regular extreme
black hole (in our case, with $n$ non-vanishing charges). Since
partially charged black holes have singular horizons of zero area,
such a process could be seen
as violating the second law of black hole mechanics/thermodynamics.
Hence the regular extreme black hole could be
expected to be classically stable. This reasoning, though, does not
apply to the singular, partially charged extreme black holes.
There also remains the possibility of quantum splitting, though it is
more unlikely.

b) Since
supersymmetry is broken, the balance of forces between constituents
is not expected to hold anymore, and they could attract each other
and compensate
the pull from the external fields. Again, this is difficult to check,
since some of the properties of the black holes, like their mass and
scalar charges, are not well defined in the presence of an
external field, and, even
worse, we can not even define the mass and
scalar charge of each individual constituent.

Even if, at present, we cannot solve this stability issue, there are
other questions to be analyzed in these black hole $+$ external
field configurations. As we will show, the shape of the black hole
happens to be distorted by the external field. From a classical,
macroscopic perspective it is not
clear whether its
area should remain unchanged. But if the area is
to be identified
with an entropy admitting a microphysical interpretation, then
we do not expect the number of microstates
of the black hole to be altered
by the introduction of the field.
This question can be unequivocally
addressed for extremal black holes
since their entropy can be expressed (in contrast to that
of non-extremal black holes)
as a function of the
conserved $U(1)$ charges alone.
Consistency with a microstate interpretation would require that
this function is not altered by the introduction of the field.
We should mention that the absence of supersymmetry
means that we can not reliably extrapolate to
the weak coupling regime where a description in terms of strings and
branes living in flat space could be possible (for the cases with
a stringy interpretation).
However, recent experience indicates that many quantities not
clearly protected by supersymmetry can be extrapolated at least in
near-supersymmetric configurations.

We have been mentioning only the case where the external forces acting
upon the black hole exactly cancel out and the black hole remains static.
But we could also have a situation
where a net force accelerates the black hole to infinity.
It is remarkable that the same
property that allows the existence of composite solutions for static
extremal black holes and Melvin fields, also makes it possible to
describe composite black holes in accelerated motion \cite{yo}.
Using these solutions we will test further the invariance of the area
(precisely, its dependence on the $U(1)$ charges) under accelerated
motions. As an outcome, we will be able to construct instantons
mediating the pair creation of composite black holes.

Another interesting feature of the theories considered above is that they
admit a large number of massless
extremal solutions. The essential point is that, due to the existence of
several kinds of gauge and scalar charges,
the BPS bound can be satisfied even when the
mass is set to zero.
In the bound state picture, zero mass is achieved by binding positive and
negative mass constituents (the latter, of course, not being allowed to
exist in isolation)
\cite{ortin}. This interpretation is often helpful,
even though the existence of the massless solutions is
independent of it and the quantum stability of such a bound state
might be problematic.

A point of concern with these {\it classical}
massless solutions of low energy string theory to qualify as solitons
is the fact that
they are nakedly singular.
Within the context of black hole solutions,
naked singularities usually arise when
a extremality condition, e.g., $M\geq |Q|$ for Reissner-Nordstrom
solutions, is violated. Now, this bound can be thought
of as a Bogomolnyi bound resulting from
partial breaking of extended supersymmetry,
and therefore one could try to exclude
nakedly singular Reissner-Nordstrom solutions
on supersymmetry grounds alone \cite{klop} (in this simple form, though,
this argument does not eliminate singular supersymmetric
Kerr-Newman solutions \cite{kalort}).
The classical singular massless solutions
we are discussing do not arise as limits
of non-extreme black holes, but
still saturate the supersymmetric BPS bound
(in the cases where a supersymmetric
embedding is known). Thus the argument could, perhaps, be
advocated that if solutions
are to be censored on the basis of supersymmetry arguments,
then the extremal
massless holes should be admitted in the theory.

Nevertheless, one could wonder whether a black hole
with Compton wavelength ($\sim 1/M$) much
larger than its Schwarzschild radius ($\sim M$) can possibly admit a
semiclassical description, so the consistency of including
massless solutions still has to be further studied.
In previous work
we have found that if classical
massless holes were allowed in the theory, then Minkowski space would,
apparently, be non-perturbatively unstable against their
production in pairs
\cite{yo}. Furthermore, the instanton mediating the tunneling process
was found to have negative action,
thus implying an exponentially enhanced pair production rate.
We will extend this result to massless holes in
the $U(1)^n$ theories.
One would be tempted to conclude
that either
some mechanism exists that forbids the pair production of massless holes,
\footnote{The question would remain, then, of how they can be created at
all, since they neither seem to appear as the result of collapse or
evaporation processes.} or, otherwise, the classical massless solutions
should be regarded as physically unacceptable.

The paper is organized as follows.
We start by reviewing in Sec.~\ref{statcomp} the solutions of
four-constituent static extreme black holes, and then introduce the more
general $n$-constituent solutions. In Sec.~\ref{melvcomp}
we develop the techniques needed to generate solutions with
Melvin field backgrounds
and analyze the properties of static black holes in the field of
a composite flux tube. More
general, accelerating solutions of C- and Ernst type
for composite black holes,
are obtained in Sec.~\ref{accecomp},
where we also construct the instantons mediating
composite black hole pair creation.
In Sec.~\ref{zeromass} we allow for negative mass constituents in order
to find massless hole solutions, and we describe the instability
of flat Minkowski space induced by these objects. In Sec.~\ref{nonext}
we present the generalization
to include non-extreme black holes, both static and
in accelerated motion. Sec.~\ref{concl}
contains some final remarks on this work.

\medskip

A word on normalization conventions. The normalization of
the gauge fields $F_{(i)}$ in the $U(1)^n$ action has been found
to conveniently simplify the formulas below, but it differs from
the `canonical' one by the prefactor $1/n$.
The charge and field parameters in our solutions
$\hat q_i$, $\widehat B_i$ are chosen
so that, for instance,
the field of a monopole is $F_{(i)}=\hat q_i \sin\theta\;
d\theta\wedge
d\varphi$, and a uniform magnetic field has $F_{(i)}^2=
2\widehat B_i^2$. This has the odd effect that
the force exerted on a charge $\hat q$ by a field
$\widehat B$ is $\hat q\widehat B/n$.
The usual form appears when considering the `canonical'
charges ${\cal Q}_i$ and fields ${\cal B}_i$, obtained as
\begin{equation}
\label{canon}
{\cal Q}_i={\hat q_i\over \sqrt{n}},
\qquad
{\cal B}_i={\widehat B_i\over\sqrt{n}}.
\end{equation}

\section{Static composite black holes}
\label{statcomp}

The extreme {\it electrically} charged black hole solutions of the
Einstein-Maxwell-scalar theory
(\ref{dilaction}) are \cite{gibmae}
\begin{eqnarray}
\label{dilbh}
ds^2 &=& -\left( 1+{\beta\over r}\right)^{-2/(1+a^2)}dt^2 +
 \left( 1+{\beta\over r}\right)^{2/(1+a^2)}\delta_{mn}dx^m dx^n,
\\
e^{-2a\phi}&=&e^{-2a\phi_0}
\left( 1+{\beta\over r}\right)^{2a^2/(1+a^2)},\qquad
A_t=-{e^{a\phi_0}\over \sqrt{1+a^2}}\left(1+{\beta\over r}\right)^{-1},
\nonumber
\end{eqnarray}
$r^2=x_m x^m$,
and the
mass $M$ and electric charge $e^{a\phi_0}Q$ given by
\begin{equation}
M={\beta\over 1+a^2},\qquad Q={\beta\over \sqrt{1+a^2}}.
\end{equation}
The asymptotic value of the scalar, $\phi_0$,
is a free parameter. In some cases we will fix it so as to match an
appropriate background at infinity.
We can also define a scalar charge, $\Sigma$,
by $\phi=\phi_0+\Sigma/r +O(r^{-2})$,
so that $\Sigma= -a \beta/(1+a^2)$, and the solution satisfies the
`antigravity' extremality condition
\begin{equation}
M^2 + \Sigma^2 = Q^2,
\end{equation}
that expresses cancellation of gravitational and scalar
attraction against electromagnetic repulsion.
This allows multi-centered black hole solutions to exist.
The {\it magnetically} charged solutions can be obtained by leaving the
canonical metric invariant and transforming
$F\rightarrow e^{-2a\phi} *F$ and $\phi
\rightarrow -\phi$, with $*F$
the Hodge dual of $F$.

The extremal black hole solutions of the theory (\ref{hetaction})
take the form
\begin{eqnarray}
\label{rahmsol}
ds^2 & = & -(\Delta_1 \Delta_2  \Delta_3  \Delta_4)^{-1}dt^2 +
 \Delta_1 \Delta_2  \Delta_3  \Delta_4 \delta_{mn}dx^m dx^n,\nonumber
\\
e^{-\eta}&=&\frac{\Delta_1\Delta_3}{\Delta_2\Delta_4}, \, \,
e^{-\sigma}=\frac{\Delta_1\Delta_4}{\Delta_2\Delta_3}, \, \,
e^{-\rho}=\frac{\Delta_1\Delta_2}{\Delta_3\Delta_4},
\\
{}F_{(1/3) \; tj} &=&\partial_j \Delta_{1/3}^{-2}, \quad
\tilde{F}_{(2/4) \; tj} =\partial_j\Delta_{2/4}^{-2},
\nonumber
\\
\Delta_i &=& \left(1+{q_i\over r}\right)^{1\over 2}\quad i=1,\dots, 4,
\nonumber
\end{eqnarray}
where $\tilde{F}_{2/4}=e^{-\eta\pm (-\sigma+\rho)}*F_{2/4}$ and the
scalars have been set to asymptote to zero at infinity.
The charges $q_1,\;q_3$ in (\ref{rahmsol}) are of electric type, whereas
$q_2,\;q_4$ are of magnetic type, but it is clear that we can apply
a duality transformation to the gauge fields and reverse the
role of electric-magnetic charges.
If all the $q_i$ are (strictly) positive, then $r=0$ is a regular
horizon of finite size,
and there is a singularity behind the horizon at $r=-\max_i\{q_i\}$.
If any
of the $q_i$ vanishes, the horizon shrinks to zero size and
the surface $r=0$ is a double null singularity.

This solution was
first found in \cite{cvetyoum2} (see also
\cite{cvetsey}) and later rediscovered in \cite{rahmfeld}.
Each of the harmonic functions $\Delta_i^2$ can be generalized so as
to give multi-center configurations.
The form of this solution suggests that black hole configurations
can be
thought of as composites of four basic building blocks, each charged
under a different Maxwell field.
In particular, the black hole
solutions of single scalar-Maxwell
theories with $a=\sqrt{3},1,1/\sqrt{3},0$ correspond to taking,
respectively, equal charges for
one, two, three or four constituent black holes sitting at the same
point, the rest of the charges
being zero (or placed at infinity).
{}For example, if we take $q_1=q_3=\sqrt{2}Q$ and
$q_2=q_4=0$, we recover the $a=1$ solution of (\ref{dilbh}), with
fields $\eta=2\phi$, $\sigma=\rho=0$,
and $F_{(1)}=F_{(3)}=\sqrt{2} F$.
Each constituent can be thought of as having a mass $m_i=q_i/4$, which is
its ADM mass when all other constituents are absent.
Except in section \ref{zeromass}, we will take
these parameters to be non-negative.
The total mass is
\begin{equation}
\label{mass}
M = {1\over 4}\sum_{i=1}^4 q_i=\sum_{i=1}^4 m_i,
\end{equation}
and therefore the binding energy is zero.

The property that allows such factorized form for the solutions is
closely linked to yet another extremal `antigravity' condition
(which is
interpreted as a supersymmetric BPS condition when the theory is
embedded in a supergravity theory). To see this, let us note, as in
\cite{rahmfeld}, that for a conformastatic ansatz
\begin{equation}
\label{conform}
ds^2=- H ^{-1}dt^2 +  H  \delta_{mn}dx^m dx^n,
\end{equation}
the Ricci tensor takes the form
\begin{eqnarray}
\label{ricci}
R_{tt} &=& -{1\over 2 H ^2}\sum_k
\partial_k^2 \log  H ,\nonumber
\\
R_{ii} &=& -{1\over 2}\sum_k \partial_k^2 \log  H
-{1\over 2}(\partial_i\log  H )^2,
\\
R_{ij} &=& -{1\over 2}\partial_i\log  H \;
\partial_j \log  H .\nonumber
\end{eqnarray}
The nonlinear terms would generate crossed terms between
the different $U(1)$ fields, e.g., $(\partial\log\Delta_1)
(\partial\log\Delta_2)$. However, it is straightforward to check
that, when the scalar fields are introduced, all the crossed terms
terms cancel out in the differential equations of motion if the
algebraic relation
\begin{equation}
\label{algebr}
(\log  H )^2+\eta^2+\sigma^2+\rho^2 =4\sum_{i=1}^4(\log \Delta_i)^2
\end{equation}
is satisfied.
{}For $H=\prod_i \Delta_i$, and defining the scalar
charges $N,\Sigma,P$ for
the scalar fields $\eta,
\sigma,\rho$ by $\eta=N/r +O(r^{-2})$, etc., then (\ref{algebr})
implies
\begin{equation}
\label{bps4}
4M^2 + N^2+\Sigma^2+P^2 =\sum_{i=1}^4 q_i^2.
\end{equation}
This is precisely the BPS bound that expresses the cancellation of
forces among constituents.

Now, given this fact that the independent action of each of the gauge
fields follows from a simple algebraic relation, we would like to
construct more general theories with this property,
such that a wider
class of dilatonic extreme black holes with coupling $a$ could be
viewed as composites. The theory defined by (\ref{unaction}) does
precisely this. Extreme black hole solutions of this theory
with electric charge under each of the $U(1)$ fields
have metric (\ref{conform}), with
\begin{equation}
\label{unbhs}
 H =\prod_{i=1}^n \Delta_i,\qquad
\Delta_i = \left( 1+{q_i\over r}\right)^{2/n},
\end{equation}
and
\begin{equation}
\label{unbhs2}
e^{-\sigma_i} =e^{-\sigma_i^0}\frac{\Delta_i^n}{ H }, \qquad
{}F_{(i) \; tj} =\pm e^{\sigma_i^0/2}\partial_j {\Delta_{i}^{-n/2}},
\end{equation}
where we have reintroduced non-vanishing scalar vevs $\sigma_i^0$,
to be of later
use, and have emphasized that the sign of the electric charges
$\hat q_i$,
\begin{equation}
\hat q_i =\pm e^{\sigma_i^0/2}\; q_i,
\end{equation}
is essentially
arbitrary.
Upon making
$F_{(i)}\rightarrow e^{-\sigma_i} *F_{(i)}$ and $\sigma_i
\rightarrow -\sigma_i$,
we can convert the electric charges into magnetic ones.
Multi-center solutions are equally possible.

The algebraic relation that
allows cancellation in the field equations
of crossed terms between different gauge fields
is
\begin{equation}
\label{algebrn}
(\log H)^2+{1\over n^2}\sum_{i>j}[(\sigma_i-\sigma_i^0)-(\sigma_j
-\sigma_j^0)]^2
=n\sum_{i=1}^n(\log \Delta_i)^2,
\end{equation}
which, in turn, can be seen as an extremality condition between
parameters,
\begin{equation}
\label{bpsbound}
M^2+{1\over 4 n^2}\sum_{i<j}(\Sigma_i-\Sigma_j)^2={1\over n}
\sum_{i=1}^n q_i^2,
\end{equation}
where $M={1\over n}\sum_{i=1}^n q_i$,
and $\Sigma_i$ is the scalar charge
associated to $\sigma_i$. The latter saturate separate bounds
of the form
\begin{equation}
\label{bound2}
{\Sigma_i \over 2}+ q_i= M.
\end{equation}

The solution (\ref{unbhs}) makes it possible to interpret,
for every rational value of $a^2$, the extreme single-Maxwell-scalar
black holes
(\ref{dilbh}) as composite objects.
To this effect, notice that if $s$ out of the $n$ possible
electric charges are equal
and non-zero,
say, $\hat q_1=\dots=\hat q_s$,
with the remaining $n-s$ charges being zero, then the black hole solution
corresponding to dilaton coupling
\begin{equation}
a=\sqrt{{n\over s}-1},
\end{equation}
is reproduced.
In fact, the fields are identified as
\begin{eqnarray}
\label{identf}
\sigma_1&=&\dots=\sigma_s=2 a\phi,\qquad \sigma_{s+1}=\dots=\sigma_n=
-{2\over a}\phi,
\\
{}F_{(i)}&=&\sqrt{a^2+1}\; F,\quad i=1,\dots,s.\nonumber
\end{eqnarray}
Therefore, we can view an extreme black hole of dilaton gravity
with $a^2=p/q$
as an $s$-hole in the $U(1)^n$ theory with $n=p+q$ and $s=q$
\footnote{One could
wonder what role is played in this construction
by the $(n-s)$ gauge fields that are set
to zero, since also an equal number of
scalar fields decouple from the theory,
and thus we are effectively left with a theory with $s$ gauge fields
and scalars.
However, the fact that
we started with $n$ fields, instead of $s$, enters through the
normalization of the scalar fields in (\ref{unaction}),
which is fixed and depends on $n$. This normalization
is actually all there is
to the coupling constant $a$, since in (\ref{dilaction}) we could
redefine $a\phi\rightarrow \phi$ so that
$a$ would only appear as fixing the normalization
of the kinetic term $(\partial\phi)^2$.}.

Notice that the `basic blocks' (i.e., the one-constituent
solutions, $s=1$) always
correspond to $a\geq 1$ (except for the trivial case $n=1$). This
provides further evidence for the conjecture
that the extreme
black holes with $a<1$ should not be viewed as elementary
particles \cite{holzhey}.

The entropy associated to the black holes in the $U(1)^4$ theory is
\cite{cvetyoum2}
\begin{equation}
\label{ent4}
S=A_{bh}/4=\pi \sqrt{q_1 q_2 q_3 q_4}=\pi \sqrt{|\hat q_1 \hat q_2
\hat q_3 \hat q_4|},
\end{equation}
whereas for a $U(1)^n$ black hole we find
\begin{equation}
\label{entn}
S=\pi \prod_{i=1}^n q_i^{2/n}=\pi\prod_{i=1}^n |\hat q_i|^{2/n}.
\end{equation}
Since the $U(1)^n$
theory does not presumably correspond, for $n>4$, to any
truncation of compactified string theory, it is uncertain whether
the expression of the entropy as a
product of charges stems from a more fundamental microscopic origin.

Notice that the entropy does not depend on the asymptotic value of
the scalars. Actually, this can be related to the elegant
principle to compute
the black hole entropy
developed by Ferrara and Kallosh, which
is based
on the existence of attractors in moduli space \cite{kalfer}.
By this it is meant that near the
black hole horizon the solution does not really depend on the
values of the moduli at infinity.
Let us explain this in a bit more detail. The extreme black hole
solution is completely determined once we specify
$2n-1$ independent parameters:
the $n-1$ independent asymptotic values of the
moduli $\sigma_i^0$ and $n$ parameters $\bar q_i$ which, for
reasons that will be immediately apparent, we choose as
$\bar q_i= \hat q_i
e^{-\sigma_i^0}$. The mass, in particular, depends on all the
parameters. However, near the horizon $r\rightarrow 0$ we find
\begin{eqnarray}
\label{attractor}
e^{-\sigma_i^0}&\rightarrow & {|\bar q_i|^2\over \prod_{j=1}^n
|\bar q_j|^{2/n}},\quad F_{(i)}\rightarrow {1\over \bar q_i}dt\wedge dr,
\\
r^2 H&\rightarrow & \prod_{j=1}^n
|\bar q_j|^{2/n},
\nonumber
\end{eqnarray}
so to fix the solution at the horizon
we need only specify the $n$ parameters $\bar q_i$, with
the values of $\sigma_i$ at infinity being arbitrary.

The entropy in this approach is found as follows. For
theories with extended supersymmetry, the mass of the extreme black
hole is equal to the modulus of a central charge of the
supersymmetry algebra,
\begin{equation}
M = |Z(\bar q_i,\sigma_i^0)|.
\end{equation}
In the present case, although we do not know whether such
a supersymmetric embedding actually
exists for arbitrary values of $n$, we will define,
according to this relation, a
quantity $|Z|$ as
\begin{equation}
\label{central}
|Z|={1\over n}\sum_{i=1}^n |\bar q_i| e^{\sigma_i^0/2}.
\end{equation}
Then, we extremize $|Z|$
as a function of the moduli $e^{\sigma_i^0}$ at fixed $\bar q_i$, taking
into account the constraint (\ref{singular}). The values of the
moduli at the fixed point thus found are
precisely given by their values at the horizon, (\ref{attractor}).
Now, following the prescription of \cite{kalfer}, the entropy of
the black hole is given in terms of the central charge at the
fixed point
as
\begin{equation}
\label{fix}
S=\pi |Z_{\rm fix}|^2.
\end{equation}
By substituting Eqs.~(\ref{central}) and (\ref{attractor})
into (\ref{fix}),
and taking into account that $\prod_{j=1}^n
|\bar q_j|=\prod_{j=1}^n
|\hat q_j|$, we find that
the entropy (\ref{entn}) is indeed reproduced.

We shall end this section by remarking that
a solution of the
$U(1)^n$ theory with $n=4$ can be converted into a solution of the
$U(1)^4$ truncation of heterotic string theory upon replacing
\begin{eqnarray}
\label{dictnry}
g_{\mu\nu}&\rightarrow & g_{\mu\nu},\nonumber
\\
F_{1/3}&\rightarrow &F_{1/3},\qquad F_{2/4}\rightarrow \tilde F_{2/4},
\\
\eta&=&{1\over 4}(\sigma_1-\sigma_2+\sigma_3-\sigma_4),\quad
\sigma={1\over 4}(\sigma_1-\sigma_2-\sigma_3+\sigma_4),\nonumber
\\
\rho&=&{1\over 4}(\sigma_1+\sigma_2-\sigma_3-\sigma_4).\nonumber
\end{eqnarray}
In some cases, though, given its special
significance for string theories
and maximal supergravity we will explicitly write the results
for the $U(1)^4$ theory (\ref{hetaction}).

\section{Black holes in composite Melvin fields}
\label{melvcomp}

We would like now to find solutions describing the composite
black holes of the previous section
as subject to (approximately uniform) external fields. In
\cite{dilatonc}, a
systematic way was presented to generate background fields
in the single-scalar-Maxwell theory (with arbitrary $a$) starting
from axisymmetric
solutions. This
involved a generalization to dilaton theories of the general relativistic
Harrison
transformation \cite{harrison}. For the single scalar, $U(1)^2$ theory,
the corresponding transformation was given in \cite{ross}. Here we
generalize the method, first to the $U(1)^4$ theory, then to $U(1)^n$.
The remarkable composite structure will be seen to appear again.

Thus, let us be given a solution of the equations of motion
of the theory in the warped product form
\begin{equation}
\label{warped}
ds^2 = g_{jk}dx^j dx^k+g_{\varphi\varphi}d\varphi^2,
\end{equation}
with $g_{jk}$ a $2+1$-dimensional metric, and gauge potentials
\begin{equation}
\label{axipot}
A_{(i)} = A_{(i)\;\varphi}d\varphi,\qquad i=1,\dots,n,
\end{equation}
where all the fields $g_{\mu\nu}$, $A_{(i)}$ and the scalars are
independent of $\varphi$. For the $U(1)^4$ theory,
the potentials $A_{(2/4)}$ are meant to be
potentials for the {\it dual} field strengths $\tilde{F}_{(2/4)}$, i.e.,
${F}_{(2/4)\;\mu\nu}={1\over 2}e^{\eta\pm (\sigma-\rho)}
\epsilon_{\mu\nu\alpha\beta}\partial^{\alpha} A_{(2/4)}^{\beta}$.

{}For ease of comparison, we quote the transformation found
in \cite{dilatonc} for the single-Maxwell-scalar theory:
\begin{eqnarray}
\label{harrisona}
g'_{jk}&=&\lambda^{2/(1+a^2)} g_{jk},\qquad
g'_{\varphi\varphi}={1\over\lambda^{2/(1+a^2)}} g_{\varphi\varphi},
\nonumber
\\
e^{-2a\phi'}&=& \lambda^{2a^2/(1+a^2)}\;e^{-2a\phi},
\\
A'_{\varphi}&=& {2\over (1+a^2)B\lambda}\left( 1+{1+a^2\over 2}
B A_{\varphi} \right),\nonumber
\\
\lambda &=& \left( 1+{1+a^2\over 2}
B A_{\varphi} \right)^2 +{1+a^2\over 4}B^2 g_{\varphi\varphi}
e^{2a\phi}.\nonumber
\end{eqnarray}
When applied to a solution of the form (\ref{warped}), (\ref{axipot}),
these transformations leave the action (\ref{dilaction}) invariant, thus
yielding a new solution. In particular, starting from flat space one
obtains the dilatonic Melvin flux tubes \cite{gibmae}.

In the same vein, we have found that new solutions
of the $U(1)^4$ theory are generated by
the transformations
\begin{eqnarray}
\label{harrison4}
g'_{jk}&=&\sqrt{\Lambda}\; g_{jk},\qquad
g'_{\varphi\varphi}={1\over\sqrt{\Lambda}} g_{\varphi\varphi},\nonumber
\\
e^{-2\eta'}&=&\frac{\lambda_1\lambda_3}{\lambda_2\lambda_4}\;e^{-2\eta},
\qquad
e^{-2\sigma'}=\frac{\lambda_1\lambda_4}{\lambda_2\lambda_3}
\;e^{-2\sigma},
\\
e^{-2\rho'}&=&\frac{\lambda_1\lambda_2}{\lambda_3\lambda_4}\;e^{-2\rho},
\nonumber
\\
A'_{(i)\;\varphi}&=& {2\over B_i\lambda_i}\left( 1+{1\over 2}
B_i A_{(i)\;\varphi} \right),\nonumber
\end{eqnarray}
where
\begin{eqnarray}
\Lambda &=&\prod_{i=1}^4\lambda_i,\nonumber
\\
\lambda_{1/3} &=& \left( 1+{1\over 2}
B_{1/3} A_{(1/3)\;\varphi} \right)^2 +{1\over 4}B^2_{1/3}
g_{\varphi\varphi}
e^{\eta \pm(\sigma+\rho)},
\\
\lambda_{2/4} &=& \left( 1+{1\over 2}
B_{2/4} A_{(2/4)\;\varphi} \right)^2 +{1\over 4}B^2_{2/4}
g_{\varphi\varphi}
e^{-\eta \pm(-\sigma+\rho)}.
\nonumber
\end{eqnarray}
The proof of invariance of the action (\ref{hetaction})
under these transformations is given in the appendix.

If we apply (\ref{harrison4}) to an asymptotically flat space, then
each of the factors
$\lambda_i$ corresponds to a background field that asymptotes to the
Melvin solution of $a=\sqrt{3}$ (Kaluza-Klein) theory.
The simplest example, in fact, is obtained by applying the transformations
to Minkowski space. Then we find
\begin{eqnarray}
\label{melvin4}
ds^2 &=& \sqrt{\lambda_1\lambda_2\lambda_3\lambda_4}
(-dt^2 + dz^2 + d\rho^2)
+{\rho^2\over \sqrt{\lambda_1\lambda_2\lambda_3\lambda_4}}
d\varphi^2,
\\
A_{(i)\;\varphi}&=&-{B_i\rho^2\over 2\lambda_i},\qquad
\lambda_i=1+{1\over 4}\rho^2 B_i^2,\quad i=1,\dots,4,\nonumber
\end{eqnarray}
where the scalars are easily found from the formulas above and the
gauge potentials have been shifted
by a gauge transformation to make them regular at the axis.
In the form given, the fields $B_1$, $B_3$ are
of magnetic type, whereas $B_2$, $B_4$ are electric.
The Melvin solutions corresponding to $B_1,B_3\neq 0$ and $B_2=B_4=0$
were also constructed in \cite{stringflux}, where, interestingly,
they have been shown to solve exactly
(i.e., to all orders in $\alpha'$) the equations of motion of the string.
The electric flux tubes $B_2$, $B_4$, are formally
related to $B_1$, $B_3$
by $S$ duality. However, due to the absence of any residual supersymmetry,
this cannot be invoked to imply exactness of the full solution.

The factorized form of the solution is immediately apparent.
In general, it can be already seen in
(\ref{harrison4}). It allows for a
composite-field interpretation of the Melvin solutions with dilaton
couplings $a=1,1/\sqrt{3},0$ in terms of $2,3,4$ `basic'
$a=\sqrt{3}$ fields
\footnote{We thank G.~Horowitz for first suggesting to us that Melvin
fields of the $U(1)^4$ theory should take a factorized form.}.
The picture is that of four parallel and concentrical flux tubes
along the $z$ axis carrying
independent magnetic or electric flux of each gauge field.

The key for a such a configuration to be possible is that the
cancellation of crossed non-linear terms in the Ricci tensor (the
latter,
of course, not being of the same form as (\ref{ricci}))
that ocurred for the extremal black holes also happens in this case,
due to an algebraic relation between
$\Lambda$ and the scalar fields analogous to (\ref{algebr}).
In this case, however, there is no  underlying supersymmetric
saturation of Bogomolnyi bounds, since the Melvin backgrounds do not
admit Killing spinors.

A direct generalization to the $U(1)^n$
theory can be naturally expected.
Indeed, we find that the Harrison transformation that
introduces magnetic
background gauge fields in an axisymmetric solution is
\begin{eqnarray}
\label{harrisonn}
g'_{jk}&=&\Lambda^{2/n} g_{jk},\qquad
g'_{\varphi\varphi}={1\over\Lambda^{2/n}}\; g_{\varphi\varphi},\nonumber
\\
e^{-\sigma_{i}'}&=&\frac{\lambda_i^2}{\Lambda^{2/n}}\;e^{-\sigma_i},
\nonumber
\\
A'_{(i)\;\varphi}&=& {2\over B_i\lambda_i}\left( 1+{1\over 2}
B_i A_{(i)\;\varphi} \right),
\\
\Lambda&=&\prod_{i=1}^n\lambda_i,\nonumber
\\
\lambda_{i} &=& \left( 1+{1\over 2}
B_{i} A_{(i)\;\varphi} \right)^2 +{1\over 4}B^2_{i} g_{\varphi\varphi}
e^{\sigma_i}.\nonumber
\end{eqnarray}
Again, the proof is deferred to the appendix.
In the particular case
where one takes $s$ of the
$B_i$ parameters and gauge fields
to be equal and sets the rest to zero, these transformations
can be seen to reproduce the dilatonic Harrison transformations
(\ref{harrisona}) for
the values $a^2=(n-s)/s$.

The solution describing a static black hole in external fields is
obtained by applying these transformations to the black hole solutions of
the previous section.
In the form given, all the fields are magnetic, whereas the charges of
the black holes in (\ref{unbhs2}) are electric. Thus, we will first
dualize the
charges of the black hole solutions so as to make them interact
non-trivially with the
external fields. The solution thus generated is
\begin{eqnarray}
\label{bhmelvinn}
ds^2 &=& \Lambda^{2/n}\left[ -{1\over H}dt^2 +H(d r^2
+ r^2 d\theta^2)\right] + {H\over\Lambda^{2/n}} r^2
\sin^2\theta d\varphi^2,\nonumber
\\
A_{(i)\;\varphi}&=& {2 e^{\sigma_i^0/2}\over
B_i\lambda_i}\left(1+{1\over 2}q_i B_i
\cos\theta\right)+k_i,
\\
e^{\sigma_i}&=&e^{\sigma_i^0}\;{\Delta_i^n\over
H} {\Lambda^{2/n}\over\lambda_i^2},\nonumber
\\
\lambda_i&=&\left(1+{1\over 2}q_i B_i
\cos\theta\right)^2 + {1\over 4}B_i^2 r^2\Delta_i^n \sin^2\theta,\nonumber
\end{eqnarray}
with $H,\;\Delta_i$ as in (\ref{unbhs}). The
asymptotic values $\sigma_i^0$ for the scalars and the constants $k_i$
fixing the position of Dirac strings
have been introduced to later match required values.

The metric (\ref{bhmelvinn}) contains, in general, conical singularities
(strings or struts) along the axes $\theta=0,\pi$.
In the case where there is a single field acting, the
singularities can not be cancelled by
any choice of parameters or period $\Delta\varphi$, the reason being that
the external force acting on the black hole cannot be balanced.
However, equilibrium can be achieved when different fields act.
Indeed, to ensure regularity along the axis we
require
\begin{equation}
\label{concanc}
\Delta\varphi=2\pi \Lambda^{2/n}|_{\theta=0}=
2\pi \Lambda^{2/n}|_{\theta=\pi},
\end{equation}
which means that
\begin{equation}
\label{statnostrut}
\prod_{i=1}^n\left( 1+{1\over 2} q_i B_i \right)=
\prod_{i=1}^n\left( 1-{1\over 2} q_i B_i \right),
\end{equation}
and for small values of $q_i B_i$ this constraint can be expressed as
\begin{equation}
\label{balance}
\sum_i q_iB_i \approx 0,
\end{equation}
i.e., the expected balance of classical forces.
We will always impose ${1\over 2} |q_i B_i|<1$; the values
$|q_i B_i|=2$ lead to singularities along any of the axes.
Eq.~(\ref{statnostrut}) requires at least two of the products
$q_i B_i$ to be non vanishing.

The parameters $q_i$ and $B_i$ are not the physical charge and field;
rather, they approximate them for small values of $q_i B_i$. The physical
charge $\hat{q}_i$ is found, as usual,
by integration of the field strengths on
a sphere surrounding the black hole, with the result
\begin{equation}
\hat{q}_i = q_i e^{\sigma_i^0/2}\;{\Lambda_0^{2/n}
\over 1-{1\over 4}q_i^2 B_i^2},
\end{equation}
where we defined $\Lambda_0\equiv \Lambda|_{\theta=0}=
\Lambda|_{\theta=\pi}$.
On the other hand, we would like to define a physical
field strength parameter
$\widehat{B}_i$
as the value of the field strength on the axis at infinity. However,
rather curiously, the latter
quantity takes different values along the axis $\theta=0$ or
$\theta=\pi$. One finds
\begin{equation}
{\widehat B_i}^2|_{\theta=0,\pi}=
{1\over 2} F_{(i)}^2|_{r\rightarrow\infty,\;\theta=0,\pi}=
B_i^2 {e^{\sigma_i^0}\over
(\lambda_i|_{\theta=0,\pi})^3},
\end{equation}
and, if $q_iB_i\neq 0$,
then
$\widehat B_i|_{\theta=0}
\neq \widehat B_i|_{\theta=\pi}$. What this means is that the
lines of force of each gauge
field get distorted due to the interaction with the black hole.
The flux of the black hole itself is asymmetrically redistributed
in the `forward-backward' directions. For
some of the gauge fields the flux lines
concentrate more in the `forward' direction, for others in the
`backward' direction, but there always remains a residual `lensing'
effect at large distances.
Notice, however,
that $\prod_i \widehat B_i|_{\theta=0}=
\prod_i \widehat B_i|_{\theta=\pi}$.

In fact, the scalar
fields also tend to different values along the axes.
As a
consequence of this, it is not possible to match the solution to an
asymptotic Melvin background simultaneously
at both semiaxes $\theta=0$ and $\theta=\pi$.
{}For definiteness we choose to match at $\theta=0$, and set
\begin{equation}
\label{choice}
e^{\sigma_i^0}={(\lambda_i|_{\theta=0})^2\over \Lambda_0^{2/n}}.
\end{equation}
Then, at large spatial distances close to the axis $\theta=0$,
the geometry becomes
\begin{eqnarray}
ds^2 &\approx& \tilde\Lambda^{2/n} (-d\bar t^2 +d\bar r^2
+\bar r^2 d\theta^2) + {\bar r^2\sin^2\theta\over \tilde\Lambda^{2/n}} d
\bar\varphi^2,\\
\tilde\Lambda &=&\prod_i\tilde\lambda_i=\prod_i \left(1+{1\over 4}
({\widehat B_i}^0)^2\bar r^2\sin^2\theta\right),\nonumber
\end{eqnarray}
with $(\bar t,\bar r)=\Lambda_0^{1/n}(t,r)$,
$\bar\varphi=\Lambda_0^{-2/n}\varphi$ and
${\widehat B_i}^0\equiv\widehat B_i|_{\theta=0}$. This is
precisely a Melvin universe with magnetic field parameter
${\widehat B_i}^0$. However,
in the
opposite direction we find
\begin{equation}
\widehat B_i|_{\theta=\pi}={\widehat B}_i^0(1+\hat q_i{\widehat B}_i^0)^3.
\end{equation}
It is interesting to find that, in terms
of the physical parameters, the no-strut condition (\ref{statnostrut})
is simply
\begin{equation}
\label{nostrutph}
\prod_{i=1}^n (1+\hat q_i{\widehat B}_i^0)=1.
\end{equation}

One can easily see that the black hole horizon is distorted due to the
external fields. The simplest way to see this is by analyzing
how the size
of the circles $\theta={\rm constant}$ changes from the equator to the
poles. A more detailed study, which can be easily carried out
for, e.g., the case with $n=2$ gauge fields,
involves computing the intrinsic
scalar curvature of the horizon.
The result in the general case is that the curvature is biggest
at the poles $\theta=0,\pi$,
and decreases monotonically to its smallest value at the equator
$\theta=\pi/2$. Furthermore, the curvature of the horizon
is always positive.
This means that the
the horizon is elongated along the axis, and always keeps a convex shape
(in particular, it never develops any `neck' at the equator).
However, even if the horizon is distorted, the area
\begin{eqnarray}
\label{area4}
A_{bh} &=& \int d\varphi\;d\theta
\sqrt{g_{\theta\theta}g_{\varphi\varphi}}|_{r=0}=4\pi\Lambda_0^{2/n}
\prod_i q_i^{2/n}\nonumber
\\
&=&4\pi\prod_i|\hat{q}_i|^{2/n}
\end{eqnarray}
(we have used the no-strut condition)
is still given by the same function (\ref{entn})
of the {\it physical} charges as in
the absence of external fields! The way in which this result appears
is rather non trivial, and it would hardly have been expected were it
not for the correspondence of the area with a number of microstates.

\section{Accelerating composite black holes}
\label{accecomp}

The analysis of the previous section has been restricted to static
configurations that satisfy the no-strut condition (\ref{nostrutph}).
In physical terms, the appearance
of conical singularities is a
rather general feature of configurations where, loosely speaking,
there is a mismatch
between the acceleration of an object and the forces acting on it. One
could say that the
conical singularities supply the forces needed to satisfy the equations
of motion (operationally, a conical defect $\delta$
can be blown up and replaced with a cosmic string
vortex with energy density $\mu=\delta/(8\pi)$).
We would like to extend our analysis of composite black holes to
situations where the black holes accelerate.
Configurations of this kind are described in general relativity by
the C-metric \cite{kinnersley} and Ernst \cite{ernst}
solutions. In
fact, they describe the
uniformly accelerated motion in opposite directions of {\it two}
oppositely
charged black holes. In the C-metric the black holes accelerate
without an external force field, and thus the geometry contains strings
or struts.
A possible way to remove them is by introducing external
gauge fields that provide the accelerating force.
Technically, this is done
by means of a Harrison transformation of the
kind discussed above, and the Ernst solution is thus obtained.

The C-metric, and its dilatonic and $U(1)^2$ generalizations given in
\cite{dilatonc,ross}, are of a rather more complicated form than the
static, spherically symmetric black hole solutions. Nevertheless, we will
still be able to construct solutions describing extremal
composite black holes in
accelerated motion. The composites are found to accelerate as a whole,
all the
constituents remaining bound. The $U(1)^4$ C-metric was already given in
\cite{yo}. The solution for arbitrary number $n$ of $U(1)$ magnetic
charges turns
out to be a rather straightforward generalization of it:
\begin{eqnarray}
\label{cmetricn}
ds^2 &=& {1\over A^2(x-y)^2}\biggl[{\cal F}(x)
\left( {1-y^2\over {\cal F}(y)}
dt^2-
{{\cal F}(y)\over 1-y^2}dy^2\right)\nonumber
\\
&+&
{\cal F}(y)\left({{\cal F}(x)\over 1-x^2}dx^2 +
{1-x^2\over {\cal F}(x)}d\varphi^2\right)\biggr],\nonumber
\\
{\cal F}(\xi) &=& \prod_{i=1}^n f_i(\xi),
\qquad f_i(\xi)= (1-q_i A\xi)^{2/n},
\\
A_{(i)\;\varphi}&=& q_i x{\sqrt{1-q_i^2 A^2}
\over f_i(x)^{n/2}},\nonumber
\\
e^{-\sigma_i} &=&
{f_i(x)^n{\cal F}(y)\over f_i(y)^n{\cal F}(x)}.\nonumber
\end{eqnarray}
The composite nature of the solution is most clearly visible from
the factorized
form of
the function ${\cal F}(y)$. Once more, this is possible due to
an algebraic relation similar to (\ref{algebrn}), and is not related to
any unbroken supersymmetry, which is absent from this solution.
The extremal
$U(1)^2$ and dilatonic
C-metrics for rational values of $a^2$
are obtained as
particular cases, but to bring them into this form
some rewriting of the parameters and
coordinates is needed. A very convenient feature of the form we use
is that all the roots of the functions involved are easily
found exactly.

A brief account of how the C-metrics
describe two black holes
accelerating apart may be helpful (see \cite{kinnersley}). We restrict
the parameters to satisfy $|q_i A|<1$ $\forall i=1,\dots, n$
\footnote{The claim in \cite{yo}
that an isometry relates the solution with $|q_i A|<1$ to that with
$|q_i A|>1$ was not correct.}.
Also, in this section all the $q_i$ will be taken to be non-negative.
Then $g_{tt}$, as a function of $y$, has three real zeros at
$y=-\infty,\;-1,\;1$ \footnote{If either none or half of the charges
vanish, there is
an additional real zero at $+\infty$, which can be identified with the
one at $-\infty$ and which, in any case, will be of no relevance
for us.}.

That the solution contains black holes that can be identified with
the static ones can be seen by realizing that the solution
(\ref{rahmsol}) appears in the limit where $A\rightarrow 0$
if we set $y\rightarrow
-1/(rA)$, $t\rightarrow At$. In fact, the zero of $g_{tt}$ at
$y=-\infty$ corresponds to the extremal horizon at $r=0$. It could be
brought to a finite coordinate distance by a simple
change of coordinates.
On the other hand, the zero
at $y=-1$ is an acceleration horizon, and the
parameter $A$ roughly measures the
acceleration of the black holes,
and also their separation when they become
closest, which is given by $2/A$. The mass of the black holes
can only be defined in an approximate way, for small accelerations, as
$M\approx{1\over n}\sum_i q_i$. Also, the coordinate patch chosen
covers only
the region of space time containing one of the black holes. The
solution can be continued to yield the remaining Rindler wedge
containing an oppositely charged black hole.

In the static limit $x$ plays the role of $\cos\theta$. In general,
when $x$ is restricted to $-1\leq x\leq 1$ and $\varphi$ is periodically
identified, the $(x,\varphi)$ sector has topology $S^2$.
The
roots $x=-1, x=1$ correspond to the poles of the sphere, and define
the axes
pointing to infinity and to the other black hole, respectively.
Asymptotic infinity is at the point $x=y$ where the conformal factor
in front of the metric diverges;
in particular, $x=y=-1$ is spacelike infinity. Hence $y$ is restricted
to $-\infty < y < x$. With coordinates and parameters within this range,
the metric has the appropriate Lorentzian signature.

The solution asymptotes to flat space, with
conical singularities along the axes $x=\pm 1$ that, for non-negative
values of $q_i$, cannot be
simultaneously cancelled everywhere. As explained, an external field
can provide the additional parameter needed to construct regular
solutions. Thus, we perform
a $U(1)^n$-Harrison transformation (\ref{harrisonn})
on (\ref{cmetricn}) to get the $U(1)^n$ Ernst solution,
\begin{eqnarray}
\label{ernstn}
ds^2 &=& {\Lambda^{2/n}\over A^2(x-y)^2}\biggl[{\cal F}(x)
\left( {1-y^2\over {\cal F}(y)}
dt^2-
{{\cal F}(y)\over 1-y^2}dy^2\right)+
{\cal F}(y){{\cal F}(x)\over 1-x^2}dx^2\biggr] \nonumber
\\
&+&
{(1-x^2){\cal F}(y)\over A^2(x-y)^2\Lambda^{2/n}
{\cal F}(x)}d\varphi^2,\nonumber
\\
{}\nonumber
\\
A_{(i)\;\varphi}&=& {2 e^{\sigma_i^0/2} \over B_i\lambda_i}\left(1+
{1\over 2} q_i B_i x{\sqrt{1-q_i^2 A^2}
\over f_i(x)^{n/2}}\right)+k_i,
\\
e^{-\sigma_i} &=& e^{-\sigma_i^0}\;
{f_i(x)^n{\cal F}(y)\over f_i(y)^n{\cal F}(x)}{\lambda_i^2 \over
\Lambda^{2/n}},\nonumber
\\
\lambda_i &=& \left(1+
{1\over 2} q_i B_i x{\sqrt{1-q_i^2 A^2} \over f_i(x)^{n/2}}\right)^2
+{1\over 4} B_i^2{1-x^2\over A^2(x-y)^2}{f_i(y)^n\over f_i(x)^n},
\nonumber
\end{eqnarray}
with ${\cal F}(\xi)$, $f_i(\xi)$ as above. Non vanishing $\sigma_i^0$'s
have been reintroduced and will be determined later.

Conical singularities in the spheres $(x,\varphi)$
can now be removed everywhere if
\begin{equation}
\Delta\varphi=2\pi \Lambda(1)^{2/n}{\cal F}(1)
=2\pi \Lambda(-1)^{2/n}{\cal F}(-1)
\end{equation}
where $\Lambda(\pm 1)\equiv\Lambda(x=\pm 1)$. This equation admits
non-trivial solutions if at least one of the products $q_i B_i$ is
non-zero. For small values of the
$q_i$ it yields, in terms of the
`canonical' fields and charges (\ref{canon}),
$MA \approx \sum_i {\cal Q}_i{\cal B}_i$,
i.e., Newton's law.

The solution can be seen to asymptote to a Melvin field at spatial
infinity $x,y\rightarrow -1$. To this effect, change to
coordinates $(\bar t,\bar y,\bar x, \bar\varphi)$ using
\begin{eqnarray}
x+1= -2L^{4/n}{\cal F}(-1)^2 {1-\bar x^2\over(\bar x+\bar y)^2},\nonumber
\\
y+1= -2L^{4/n}{\cal F}(-1)^2 {1-\bar y^2\over(\bar x+\bar y)^2},
\\
t={\cal F}(-1)\;\bar t,\quad \varphi= {\cal F}(-1) L^{4/n}\;\bar\varphi,
\nonumber
\end{eqnarray}
where $L^2\equiv \Lambda(x=-1)$. For $x,y\rightarrow -1$ the
$U(1)^n$-Ernst metric then goes to
\begin{eqnarray}
ds^2 &\approx& {\tilde\Lambda^{2/n}\over A^2 (\bar x-\bar y)^2}
\left[ (1-\bar y^2)d\bar t^2-{d\bar y^2\over 1-\bar y^2} +
{d\bar x^2\over 1-\bar x^2}\right]
\\
&+&{1-\bar x^2\over A^2(\bar x-\bar y)^2
\tilde\Lambda^{2/n}}d\bar\varphi^2,
\nonumber
\end{eqnarray}
with
\begin{eqnarray}
\label{ernstasym}
\widehat B_i&=&{B_i\over \ell_i L^{2/n}{\cal F}(-1)},\qquad
\ell_i^2 =\lambda_i(x=-1)
\\
\tilde\Lambda&=&\prod_{i=1}^n\left( 1+{1\over 4}
\widehat B_i^2{1-\bar x^2\over A^2(\bar x-\bar y)^2}\right).
\nonumber
\end{eqnarray}
This is a Melvin background with magnetic field parameter $\widehat B_i$,
though
written in somewhat unusual accelerated
coordinates \cite{extremec}. The scalar and gauge fields can be
matched if we set
\begin{equation}
e^{\sigma_i^0}={\ell_i^4\over L^{4/n}}.
\end{equation}
With this value, the magnetic field on the
axis $x=-1$ at large spatial distances, $y\rightarrow -1$,
\begin{equation}
{1\over 2} F_{(i)}^2|_{x=-1}\rightarrow B_i^2 {e^{\sigma_i^0}\over
\ell_i^6 {\cal F}(-1)^2},
\end{equation}
is consistent with the physical field parameter in (\ref{ernstasym}).

As of the physical charges of the black holes in the
$U(1)^n$-Ernst metric, they are readily computed as
\begin{equation}
\hat q_i =q_i\;{{\cal F}(-1) L^{2/n} \ell_i\over \sqrt{1-q_i^2 A^2}
\sqrt{\lambda_i(x=1)}}.
\end{equation}

We have already found that external fields do not change the area (i.e.,
the entropy)
of the extreme black holes. Does the acceleration affect it?
The answer,
\begin{eqnarray}
A_{bh} &=& \int d\varphi\;dx
\sqrt{g_{xx}g_{\varphi\varphi}}|_{y=-\infty}=
4\pi\Lambda_0^{2/n}{\cal F}(1)
\prod_i q_i^{2/n}\nonumber
\\
&=&4\pi\prod_i |\hat{q}_i|^{2/n},
\end{eqnarray}
once again, shows that the dependence on the physical charges
remains unaltered.

The shape of the black holes can be studied as we
did in the static case, but the formulas get rather complicated.
The result is that, in the general situaton where there are at
least two non-equal gauge fields,
the black holes are again elongated
along the axis (this was also noticed in \cite{ross}). Curiously, when
there is effectively only one gauge field excited, i.e., when all the
non-vanishing gauge fields are equal, the solution near the horizon
approaches exactly the spherically symmetric
static black hole solution \cite{extremec}.
At present it is not clear to us why this should be so.

The Lorentzian solutions just described can be easily continued to
imaginary time $\tau=it$. When $y$ has range $-\infty <y \leq -1$ and
$\tau$ is periodically identified to satisfy regularity at the
acceleration horizon, i.e.,
\begin{equation}
\Delta\tau= 2\pi {\cal F}(-1),
\end{equation}
we get an exact Euclidean instanton with topology $S^2\times{\bf R}^2-
\{ pt\}$
mediating the pair production
of these extreme black holes. For the particular case of
the $U(1)^4$ theory,
this includes, e.g.,
the pair creation of Kaluza-Klein monopoles \cite{extremec}
and $H$-monopoles.

The pair creation rate is calculated,
in the leading approximation, as $\Gamma \sim e^{-I}$,
with $I$ the classical Euclidean
action of the instanton.
This quantity is given by the usual boundary terms at infinity,
after having matched
the boundary geometry and fields to those of the Melvin background
\cite{hhr}.
The Hamiltonian vanishes, and the action
is fully given by the difference between the areas of the
acceleration horizons in the $U(1)^n$ Ernst metric and in the reference
Melvin
space. The final exact answer takes the form
\begin{equation}
\label{ernstactn}
I= -{1\over 4}\Delta A_{acc}={\pi L^{4/n} {\cal F}(-1)^2\over n A}
\sum_{i=1}^n {q_i\over 1+q_i A}.
\end{equation}
This result can be seen to agree with those previously found for the
particular cases considered in
\cite{extremec,ross}. For small charges,
and using the no strut condition, one finds
\begin{equation}
\label{schwinger}
I\approx {\pi M^2\over \sum_{i=1}^n {\cal Q}_i {\cal B}_i},
\end{equation}
i.e., the generalized Schwinger value $I\approx\pi M^2/F$,
with $F$ the driving force.

We should mention that, as in \cite{hhr},  we
have not included the contribution
from the extreme black hole area in the calculation
of the instanton action above. This term, if added in the cases where
the black hole horizon is regular, would yield
an enhancement of
pair creation rates due to macroscopic indistinguishability of
the internal
states of the black holes created. Since the extremal black
holes, at least the stringy ones, appear to have non-zero entropy,
the term $-A_{bh}/4$ should presumably be added to (\ref{ernstactn})
to find the pair creation rate.

Notice that the sign of the instanton action (\ref{ernstactn})
is controlled by the term $\sum_i q_i/(1+q_i A)$ (recall that we
only allow $|q_iA|<1$ and therefore ${\cal F}(\pm 1)>0$). This implies
that $I$ is always positive for non-negative $q_i$. In next
section we will discuss a different situation.

\section{Massless holes}
\label{zeromass}

Zero-mass extreme black holes have recently attracted
interest. We would like, however, to distinguish
between two different objects, both referred to as massless black holes,
that have been described
in the literature and whose relationship is not clear.
Strominger showed in \cite{andy} that certain singularities (conifold
points) of
the moduli space of string vacua could be understood as the effect of
extremal black holes becoming massless.	This is very similar to the
mechanism by which phase transitions in $N=2$ Yang-Mills are triggered
by condensation of massless BPS monopoles. It must be stressed, though,
that since the Lagrangian is singular at the conifold,
these massless black holes do not admit a
semiclassical description.

The other class of massless black holes corresponds to
nakedly singular solutions of a regular classical Lagrangian
\cite{behrndt,kallosh,cvetyoum}.
It is not clear whether these solutions
can be consistently
identified with the massless solitons of \cite{andy}.
They preserve
a fraction of the supersymmetries of the theory, but, on account of
their naked singularities, they also present a number of bizarre
properties.
These holes, though massless,
are found to follow timelike trajectories, and in fact they can
be at rest. Since their effect on test
particles is repulsive \cite{kallosh}
we will avoid to refer to them as `black'.

Single-scalar-Maxwell theories do not have
non-trivial extreme massless solutions: the extreme black holes in
these theories are
essentially determined by only one parameter, and the mass can not be
set
to zero without also setting to zero the other charges.
As was argued in \cite{ortin}, massless holes can be easily found in
theories admitting composite black holes. All one has to do is to allow
for negative `constituent masses', something not evidently
inconsistent as long as the total mass of
the composite is non-negative. Thus, for example, in the $U(1)^4$ theory
we can choose $q_1=-q_3>0$ and $q_2=q_4=0$ to get a solution with zero
ADM mass. In this case, a naked curvature singularity appears at
$r=-q_3$, but the Bogomolnyi bound can still be non-trivially saturated.
The same thing applies to the $U(1)^n$ theories, which admit a
large variety of such
solutions for any $n\geq 2$, with a naked singularity
at $r=\max_i\{ -q_i\}$.

We have not found any surprises in the analysis of static massless
solutions subject to external fields.
However, the accelerating solutions do indeed present
a striking behavior. For the $U(1)^4$ theory they were studied in
\cite{yo}. Here we will extend the results to $U(1)^n$ theories.

Consider the $U(1)^n$ C-metric (\ref{cmetricn}).
No external field acts upon the
black hole, and the solution asymptotes to Minkowski space.
The no-strut condition
reads
\begin{equation}
\Delta\varphi =2\pi {\cal F}(1)=2\pi {\cal F}(-1).
\end{equation}
This equation can be non-trivially solved only
if negative values of the $q_i$ are
allowed. In fact, the solution for arbitrary values of $A$ requires
\begin{eqnarray}
\label{zero}
\sum_{i=1}^n q_i &=& 0,\\
\sum_{i<j<k} q_i q_j q_k &=& 0,\nonumber\\
&\vdots&\nonumber\\
\sum_{i_1<i_2<\dots<i_r} q_{i_1}\dots q_{i_r}&=& 0 \quad
{\rm for\ all\ odd\ } r\leq n.\nonumber
\end{eqnarray}
In particular, the first equation (\ref{zero}) implies that
the hole has zero mass \footnote{Or, rather, since the mass is not
defined in any precise way for the accelerating object, we should say
that there is a one-to-one
correspondence with the static massless solution in the limit
$A\rightarrow 0$.}. Hence, objects satisfying these conditions
accelerate freely and uniformly
in the otherwise empty Minkowski space.
Notice that it is not clear at all how to change the state
of motion of a particle that is following a timelike
trajectory but nevertheless has zero mass. This is another indication
that, although they are
classical solutions,
the massless holes exhibit a behavior markedly different from that of
any classical particle.

Since some of the $q_i$ are negative, these solutions have a naked
singularity at $y=\max_i\{(q_iA)^{-1}\}$,
equivalent to the singularity of the massless
static solution at $r=\max_i\{-q_i\}$. If the latter is to be regarded as
acceptable, then
we would expect the Euclidean
solution obtained from the C-metric to be a valid instanton
mediating the pair creation of these massless objects. This would
describe a putative decay of Minkowski space.

The Euclidean action of the instanton
can be readily obtained from (\ref{ernstactn})
by setting the fields $B_i$
to zero and enforcing the conditions (\ref{zero}).
The result, that generalizes the one in
\cite{yo}, can be given in a simple, exact form as
\begin{equation}
\label{decay}
I=-{\pi \over n}\sum_{i=1}^n \hat{q}_i^2,
\end{equation}
i.e., it is essentially given by the mean square physical charge.
{\it This action is
negative}. Such result could not have been possible had not we
allowed for naked singularities. It seems to imply the undesired
conclusion that Minkowski
space would be wildly unstable against pair creation of classical
massless particles. Apparently, then, either an argument ruling out
the instanton, but not the solutions with static
naked singularities, is found, or otherwise
the massless solutions should be deemed as unphysical\footnote{Some
possible objections to this line of reason were
considered in \cite{yo}.}.

Similarly, a Melvin background could also decay by forming pairs of
massless holes. The action, which can also be read
from (\ref{ernstactn}),
is again given
(but now only to leading order for small charges)
by the same negative value
(\ref{decay}). This decay mode, then, would dominate by far over the
creation of massive black holes.

We conclude this section by stressing that the purely
quantum massless black holes of \cite{andy} are not
affected in principle by this result. We can not use a
classical action to
construct a decay instanton, since the Lagrangian
becomes singular when the massless
black holes appear. The same thing could be said of massless non-abelian
monopoles if we tried to analyze their pair creation
along the lines of \cite{manton}.

\section{Non-extremal static and accelerated black holes}
\label{nonext}

Thus far we have been concentrating only on the solutions describing
extreme black holes. These are the ones that allow for a proper composite
interpretation, and their entropy can be defined solely in terms of
conserved charges. Nevertheless, it is possible to construct the
non-extreme versions of these black holes
for the general $U(1)^n$ theory
in a rather straightforward way. Remarkably enough, the corresponding
non-extremal C-metrics and Ernst solutions
can also be easily found.

Let us start by writing the non-extreme dilaton black hole solution
of (\ref{dilaction}) in the
following convenient way
\begin{eqnarray}
\label{nextdil}
ds^2&=&-{1-{r_0/ r}\over \left(1+{\beta/ r}\right)^{2/(1+a^2)}} dt^2
+{\left(1+{\beta/ r}\right)^{2/(1+a^2)}\over 1-{r_0/ r}}dr^2
+r^2 \left(1+{\beta\over r}\right)^{2/(1+a^2)}d\Omega^2,
\\
e^{-2a\phi}&=&
\left( 1+{\beta\over r}\right)^{2a^2/(1+a^2)},\qquad
A_t=-{Q\over \beta}\left(1+{\beta\over r}\right)^{-1},\nonumber
\end{eqnarray}
where we have set $\phi_0=0$, and the electric charge is
\begin{equation}
Q=\beta\sqrt{1+r_0/\beta\over 1+a^2}.
\end{equation}
The outer horizon is at $r=r_0$, and the inner horizon at $r=0$. The
parameter $r_0$ is equal to zero at extremality.

We observe that this
solution can be obtained from the extreme one by essentially introducing
the factor $1-r_0/r$ in $g_{tt}$ and $g_{rr}$. Is this feature
generalizable
to the $U(1)^n$ theory? We find that the answer is yes. The simple
modification
\begin{eqnarray}
ds^2 &=& -{1-{r_0/ r}\over H}dt^2 + {H \over 1-{r_0/ r}}dr^2
+r^2 Hd\Omega^2,
\\
A_{(i)\;t}&=&{\sqrt{1+r_0/q_i}\over \Delta_i^{n/2}},\nonumber
\end{eqnarray}
with $H$, $\Delta_i$ and the scalar fields as in the extreme solution
(\ref{unbhs}), (\ref{unbhs2}), turns out to be the correct solution.
The electric charges are in this case
\begin{equation}
\hat q_i=q_i\sqrt{1+{r_0\over q_i}},
\end{equation}
and the mass
\begin{equation}
M={r_0\over 2}+{1\over n}\sum_{i=1}^n q_i={1\over n}\sum_{i=1}^n
\left(\hat q_i^2 + {r_0^2\over 4}\right)^{1/2}
\end{equation}
does not saturate any longer the extremal bound.
For four $U(1)$ charges this solution was
previously found in \cite{nextcy}.
By taking $s$ equal nonvanishing charges, with the remaining ones being
zero,
we reproduce, with
the same identifications as in (\ref{identf}),
the non-extreme dilaton black holes (\ref{nextdil}).

The horizon area of these black holes,
\begin{equation}
A_{bh}=4\pi\prod_{i=1}^n (q_i+r_0)^{2/n}
=4\pi\prod_{i=1}^n \left[\left(\hat q_i^2+{r_0^2\over 4}\right)^{1/2}
+{r_0\over 2}\right]^{2/n},
\end{equation}
depends on $r_0$ and cannot be expressed in terms of conserved charges
only.

With little extra effort we can find non-extreme C-metrics. Once again,
the simplest possibility turns out to yield the correct answer,
\begin{eqnarray}
\label{nonextc}
ds^2 &=& {1\over A^2(x-y)^2}\biggl[{\cal F}(x)
\left( {(1-y^2)(1+r_0 Ay)\over {\cal F}(y)}
dt^2-
{{\cal F}(y)\over (1-y^2)(1+r_0 Ay)}dy^2\right)\nonumber
\\
&+&
{\cal F}(y)\left({{\cal F}(x)\over (1-x^2)(1+r_0 Ax)}dx^2 +
{(1-x^2)(1+r_0 Ax)\over {\cal F}(x)}d\varphi^2\right)\biggr],
\\
{}\nonumber
\\
A_{(i)\;\varphi}&=& q_i x{\sqrt{(1+r_0/q_i)(1-q_i^2 A^2)}
\over f_i(x)^{n/2}}.\nonumber
\end{eqnarray}
$\cal F$, $f_i$
and the scalars are as in (\ref{cmetricn}).
In the form given, the black holes have magnetic charge. Again, to bring
this solution into the form of the particular cases
considered in \cite{dilatonc,ross}, the coordinates and parameters
have to be somewhat transformed.

{}For non-negative $q_i$'s, we now
find four real zeroes of $g_{tt}$
at $y=-\infty, \;-1/(r_0 A),\;
-1,\; +1$.
With $0\leq r_0 A<1$ and $-\infty <y<x$, $-1\leq x\leq 1$, the
solution describes a non-extreme black hole and its antiparticle
accelerating in opposite directions; $y=-\infty, -1/(r_0 A), -1$ are,
respectively, the inner horizon and outer horizon of the black hole,
and the acceleration horizon.

It is now easy to perform a $U(1)^n$ Harrison transformation
on any of these solutions, and thus introduce background Melvin
fields. In particular, in this way we obtain non-extreme $U(1)^n$
Ernst solutions,
that can be continued to imaginary time so as to find instantons for
non-extreme black hole pair creation. Euclidean regularity
requires matching the temperature of black hole and acceleration
horizons.
Since the analysis is rather
straightforward, we will not give further details,
and shall only quote
the value of the instanton action mediating non-extreme black hole
pair production.
After imposing the instanton regularity conditions, and adding
the contribution from the black hole area,
the exact result is found to be
\begin{equation}
I= -{1\over 4}(\Delta A_{acc} +A_{bh})={\pi L^{4/n}
{\cal F}(-1)^2\over n A (1-r_0 A)}
\sum_{i=1}^n {q_i\over 1+q_i A},
\end{equation}
with
\begin{equation}
L=\prod_{i=1}^n\left( 1-{1\over 2} q_i B_i
\sqrt{(1+r_0/q_i)(1-q_i A)\over 1+q_i A}\right).
\end{equation}
Again, this can be readily seen to agree, for small $r_0,\;q_i$, with
Schwinger's result (\ref{schwinger}).

\section{Concluding remarks}
\label{concl}

We would like to highlight two results from this
analysis of composite black holes in external fields.
First, the invariance of the extremal
black hole entropy as expressed in terms of conserved charges alone,
and, second, the fact that the compositeness property
is not restricted to
solutions preserving some supersymmetry, but also to backgrounds
that do not admit Killing spinors, like Melvin flux tubes or C-metrics.
The compositeness seems to be better related to algebraic
relations of the type of
(\ref{algebrn}), which only for some classes of solutions (static
extreme black holes) coincide with saturated Bogomolnyi bounds.
This algebraic property, in particular, makes it possible to
construct solutions
containing an arbitrary number of parameters even for cases as
complex as the Ernst solutions.

By introducing an arbitrary number of gauge fields we have been able to
extend the composite interpretation to all the dilatonic black holes
with
rational value of $a^2$. One may wonder
whether this cannot be extended to all real values of
$a$. Formally, this can be done by allowing for an {\it infinite}
number of gauge fields. Though the consistency of such a theory might
be problematic, it is amusing to see how this would work. Consider, then,
replacing the discrete index $i$ in the action (\ref{unaction}) with
a continuous parameter $\xi$ with range $0\leq \xi\leq 1$, and let the
fields $\sigma(\xi)$, $F_{\mu\nu}(\xi)$ depend on it. The new action is
\begin{equation}
\label{uxaction}
I={1\over 16\pi G}\int d^4x\sqrt{-g}\biggl\{ R-{1\over 2}
\int_{0}^1 d\xi_1\int_{\xi_1}^1
d\xi_2\left[\partial\sigma(\xi_1)-
\partial\sigma(\xi_2)\right]^2
-\int_{0}^{1} d\xi\;
e^{-\sigma(\xi)} F(\xi)^2\biggr\},
\end{equation}
with, now, $\int_{0}^{1} d\xi\;\sigma(\xi)=0$.
It is curious that, if we turn off the gravity and scalar interactions,
the Maxwell field action can be seen as defined on a five dimensional
space $M_4\times I$, with $I$ the unit
interval, and $F_{\mu\xi}=0$. This interpretation, though, does not
seem to be generalizable to the whole action (\ref{uxaction}).

Extreme black hole solutions of this theory are given in terms of a
function $q(\xi)$ which we take to be non-negative. The metric is of
the same form (\ref{conform}), but now with
\begin{equation}
\log H=2\int_{0}^{1} d\xi\;\log\left(1+{q(\xi)\over r}\right)
\end{equation}
and
\begin{equation}
e^{-\sigma(\xi)}={\left(1+{q(\xi)/ r}\right)^2\over H},\qquad
A_t(\xi)=\left(1+{q(\xi)\over r}\right)^{-1}.
\end{equation}
The mass is $M=\int d\xi\;q(\xi)$, and the horizon area
\begin{equation}
A_{bh}=4\pi\exp\left(2\int_{0}^{1} d\xi\;\log q(\xi)\right).
\end{equation}
To reproduce the dilaton black holes with generic coupling $a$,
take $q(\xi)$ to be a step function,
\begin{eqnarray}
q(\xi) &=& q\quad {\rm for}\,\, 0\leq x <{1\over 1+a^2},\nonumber
\\
&=& 0\quad {\rm for}\,\, {1\over 1+a^2}< x \leq 1.\nonumber
\end{eqnarray}
The formal generalization of all other results in the paper is
straightforward.

A different `origin' for extreme dilatonic black holes with
$a=\sqrt{p/(p+2)}$ has been given in \cite{ght}, where, for $p$ odd,
they are found to arise in a non-singular way from non-dilatonic
$p$ branes in $(4+p)$ dimensions. It
would certainly be interesting to find the connection between their
approach and the one in this paper. In a similar vein, we believe that
generalizations
of the $U(1)^n$ action (\ref{unaction})
should also exist at least for five dimensions. Although
C-metrics are known essentially only for four dimensional
black holes, Melvin
flux tubes are more amenable to generalizations to higher dimensions.
Thus, it appears that the analysis carried out in Sec.~\ref{melvcomp}
could be extended to higher dimensional black holes (specifically,
five dimensional), and extended objects like $p$-branes
(for related extensions, see \cite{dggh}). Results similar
to those presented in this paper are expected to hold.

\section*{Acknowledgements}

We would like to thank G.~Horowitz for suggestions and for comments
on the manuscript, and P.~M.~Llatas
for conversations on these issues.
This work has been partially supported by FPI (MEC-Spain) program,
and by CICYT AEN-93-1435 and UPV 063.310-EB225/95.


\appendix

\section{$U(1)^n$ Harrison transformations}\label{apa}

In this appendix we outline the proof that the generalized
Harrison transformations (\ref{harrison4}) and (\ref{harrisonn}) leave
the actions (\ref{hetaction}) and (\ref{unaction}), respectively,
invariant.

In fact,
the proof proceeds by direct check. However, the algebra
can be simplified by rewriting the action in an
appropriate manner. Since the truncated heterotic $U(1)^4$ action can be
mapped into the $U(1)^n$ action by identifying fields as in
(\ref{dictnry}), we will only present the result for
the latter case.

Given an axisymmetric solution of the form (\ref{warped}), we denote
\begin{equation}
g_{\varphi\varphi}=V,\qquad g_{jk}={}^3 g_{jk},
\end{equation}
and then integrate the cyclic coordinate $\varphi$ in the action.
We obtain the following effective
three-dimensional action
\begin{equation}
\label{redaction}
I={\Delta\varphi\over 16\pi G}\int d^3x\sqrt{-^3g}V^{1/2}
\biggl\{ {}^3R
-{1\over 2 n^2}
\sum_{i<j}(\partial\sigma_i-\partial\sigma_j)^2 -
{2\over n V}\sum_{i=1}^n
e^{-\sigma_i} (\partial A_{(i)\varphi})^2\biggr\}.
\end{equation}
The transformations
(\ref{harrisonn}) act conformally on the metric ${}^3 g_{jk}$.
For the $U(1)^4$ action (\ref{hetaction}) one might also consider writing
things in terms of the conformally related
string metric $d\bar s^2=
e^{\eta}ds^2$, but due to the presence of other scalar fields this turns
out to be
not very advantageous.
Lengthy but straightforward algebra shows now that the
action (\ref{redaction}) is left invariant under these transformations.


\begin{thebibliography}{99}

\bibitem{it} A. Strominger and C. Vafa, {\sl Microscopic origin of
the Bekenstein-Hawking entropy}, Phys. Lett. {\bf B379} (1996) 99
(hep-th/9601029); for a review and references, see
G.T. Horowitz, {\sl The origin of black hole entropy in
string theory} (gr-qc/9604051).

\bibitem{cvetyoum2} M. Cvetic and D. Youm, {\sl Dyonic BPS saturated
black holes of heterotic string on a six-torus}, Phys. Rev. D{\bf 53}
(1996) 584 (hep-th/9507090).

\bibitem{cvetsey} M. Cvetic and A. Tseytlin, {\sl Solitonic strings and
BPS saturated dyonic black holes}, Phys. Rev. D{\bf 53} (1996) 5619
(hep-th/9512031).

\bibitem{rahmfeld} J. Rahmfeld, {\sl Extremal black holes as bound states},
Phys. Lett. {\bf B372} (1996) 198 (hep-th/9512089).

\bibitem{comptse} A. A. Tseytlin, {\sl Composite black holes in
string theory}, (gr-qc/9608044).

\bibitem{gibmae} G. W. Gibbons and K. Maeda, {\sl Black holes and
membranes in higher-dimensional theories with dilaton fields},
Nucl. Phys. {\bf B298} (1988) 741.

\bibitem{klop} R. Kallosh, A. Linde, T. Ort{\'\i}n, A. Peet
and A. Van Proeyen, {\sl
Supersymmetry as a cosmic censor}, Phys. Rev. D{\bf 46} (1992) 5278
(hep-th/9205027).

\bibitem{duffrahm2} M. J. Duff and J. Rahmfeld, {\sl Bound states of
black holes and other $p$-branes} (hep-th/9605085).

\bibitem{lupope} H. L\"u and C. N. Pope, {\sl $p$-brane solitons in
maximal supergravities},  Nucl. Phys. {\bf B465} (1996) 127
(hep-th/9512012);
{\sl Multi-scalar $p$-brane solitons} (hep-th/9512153).

\bibitem{melvin} M. A. Melvin, {\sl Pure electric and magnetic geons},
Phys. Lett. {\bf 8} (1964) 65.

\bibitem{yo} R. Emparan, {\sl Massless black hole pairs in string theory},
Phys. Lett. {\bf B387} (1996) 721 (hep-th/9607102).

\bibitem{ortin} T. Ort{\'\i}n, {\sl Massless string theory black holes
as black diholes and quadruholes}, Phys. Rev. Lett. {\bf 76} (1996) 3890
(hep-th/9602067).

\bibitem{kalort} R. Kallosh and T. Ort{\'\i}n, {\sl
Supersymmetry, trace anomaly, and naked singularities}, (hep-th/9404006).

\bibitem{holzhey} C. Holzhey and F. Wilczek,
{\sl Black holes as elementary particles}, Nucl. Phys. {\bf B380}
(1992) 447 (hep-th/9202014).

\bibitem{kalfer} S. Ferrara and R. Kallosh, {\sl Supersymmetry and
attractors}, Phys. Rev. D{\bf 54} (1996) 1514 (hep-th/9602136).

\bibitem{stringflux} A. A. Tseytlin, {\sl Melvin solution in string
theory}, Phys. Lett. {\bf B346} (1995) 55 (hep-th/9411198);
J. G. Russo and A. A. Tseytlin,
{\sl Exactly solvable string models of curved space-time backgrounds},
Nucl. Phys. {\bf B449} (1995) 91 (hep-th/9502038);
{\sl Magnetic flux tube models in superstring theory},
Nucl. Phys. {\bf B461} (1996) 131 (hep-th/9508068).

\bibitem{kinnersley} W. Kinnersley and M. Walker, {\sl Uniformly
accelerating charged mass in General Relativity}, Phys. Rev. D{\bf 2}
(1970) 1359.

\bibitem{ernst} F. J. Ernst, {\sl Removal of the nodal singularity
of the C-metric}, J. Math. Phys. {\bf 17} (1976) 515.

\bibitem{harrison} B. Harrison, J. Math. Phys. {\bf 9} (1968) 1744.

\bibitem{dilatonc} F. Dowker, J.P. Gauntlett, D. Kastor and J. Traschen,
{\sl Pair creation of dilaton black holes},
Phys. Rev. D{\bf 49} (1994) 2909 (hep-th/9309075).

\bibitem{ross} S.F. Ross, {\sl Pair production of black holes in a
$U(1)\otimes U(1)$ theory}, Phys. Rev. D{\bf 49} (1994) 6599
(hep-th/9401131);
{\sl Pair creation rate for $U(1)^2$
black holes}, ibid. D{\bf 52} (1995) 7089 (gr-qc/9509010).

\bibitem{extremec} F. Dowker, J.P. Gauntlett,
S.B. Giddings, and G.T. Horowitz, {\sl Pair creation of extremal black
holes and Kaluza-Klein monopoles}, Phys. Rev. D{\bf 50} (1994) 2662
(hep-th/9312172).

\bibitem{hhr} S. W. Hawking, G. T. Horowitz and S. F. Ross, {\sl
Entropy, area, and black hole pairs}, Phys. Rev. D{\bf 51} (1995) 4302
(gr-qc/9409013).

\bibitem{andy} A. Strominger, {\sl Massless black holes and conifolds in
string theory}, Nucl. Phys. {\bf B451} (1995) 96 (hep-th/9504090).

\bibitem{behrndt} K. Behrndt, {\sl About a class of exact string
backgrounds}, Nucl. Phys. {\bf B455} (1995) 188 (hep-th/9506106).

\bibitem{kallosh} R. Kallosh and A. Linde, {\sl Exact supersymmetric
massive and massless white holes}, Phys. Rev. D{\bf 52} (1995) 7137
(hep-th/9507022); {\sl Supersymmetric balance
of forces and condensation of BPS states},  Phys. Rev. D{\bf 53}
(1996) 5734 (hep-th/9511115).

\bibitem{cvetyoum} M. Cvetic and D. Youm, {\sl Singular BPS saturated
states and enhanced symmetries of four-dimensional N=4 supersymmetric
string vacua}, Phys. Lett. {\bf B359} (1995) 87 (hep-th/9507160).

\bibitem{manton} I. K. Affleck and N. S. Manton, {\sl Monopole pair
production in a magnetic field}, Nucl. Phys. {\bf B194} (1982) 38.

\bibitem{nextcy} M. Cvetic and D. Youm, {\sl BPS saturated and
non-extreme states in abelian Kaluza-Klein theory and effective
$N=4$ supersymmetric string vacua}, (hep-th/9508058).

\bibitem{ght} G. W. Gibbons, G. T. Horowitz, and P. K. Townsend,
{\sl Higher dimensional resolution of dilatonic black hole
singularities}, Class. Quant. Grav. {\bf 12} (1995) 297
(hep-th/9410073).

\bibitem{dggh} H. F. Dowker, J. P. Gauntlett, G. W. Gibbons, and
G. T. Horowitz, {\sl Nucleation of $p$-branes and fundamental strings},
Phys. Rev. D{\bf 53} (1996) 7115 (hep-th/9512154).


\end{thebibliography}
\end{document}